\documentclass[sigconf]{acmart}

\AtBeginDocument{%
  \providecommand\BibTeX{{%
    \normalfont B\kern-0.5em{\scshape i\kern-0.25em b}\kern-0.8em\TeX}}}

\copyrightyear{2022}
\acmYear{2022}
\setcopyright{acmcopyright}\acmConference[ASE '22]{37th IEEE/ACM International Conference on Automated Software Engineering}{October 10--14, 2022}{Rochester, MI, USA}
\acmBooktitle{37th IEEE/ACM International Conference on Automated Software Engineering (ASE '22), October 10--14, 2022, Rochester, MI, USA}
\acmPrice{15.00}
\acmDOI{10.1145/3551349.3556969}
\acmISBN{978-1-4503-9475-8/22/10}
\usepackage{enumitem}   
\usepackage{multirow}
\usepackage{float}
\def\anote#1{#1}
\begin{document}

\title{An Empirical Study of Automation in Software Security Patch Management}
\author{Nesara Dissanayake}
\affiliation{%
  \institution{CREST - Centre for Research on Engineering Software Technologies, The University of Adelaide}
  \city{Adelaide}
  \country{Australia}}
\email{nesara.madugodasdissanayakege@adelaide.edu.au}

\author{Asangi Jayatilaka}
\affiliation{%
  \institution{CREST - Centre for Research on Engineering Software Technologies, The University of Adelaide}
  \city{Adelaide}
  \country{Australia}}
\email{asangi.jayatilaka@adelaide.edu.au}

\author{Mansooreh Zahedi}
\affiliation{%
  \institution{University of Melbourne}
  \city{Melbourne}
  \country{Australia}}
\email{mansooreh.zahedi@unimelb.edu.au}

\author{Muhammad Ali Babar}
\affiliation{%
  \institution{CREST - Centre for Research on Engineering Software Technologies, The University of Adelaide}
  \city{Adelaide}
  \country{Australia}}
\email{ali.babar@adelaide.edu.au}

\renewcommand{\shortauthors}{Dissanayake, et al.}

\begin{abstract}


Several studies have shown that automated support for different activities of the security patch management process has great potential for reducing delays in installing security patches. However, it is also important to understand how automation is used in practice, its limitations in meeting real-world needs and what practitioners really need, an area that has not been empirically investigated in the existing software engineering literature. This paper reports an empirical study aimed at investigating different aspects of automation for security patch management using semi-structured interviews with 17 practitioners from three different organisations in the healthcare domain. The findings are focused on the role of automation in security patch management for providing insights into the as-is state of automation in practice, the limitations of current automation, how automation support can be enhanced to effectively meet practitioners' needs, and the role of the human in an automated process. Based on the findings, we have derived a set of recommendations for directing future efforts aimed at developing automated support for security patch management.

\end{abstract}

\ccsdesc[500]{Security and privacy~Software security engineering}
\ccsdesc[500]{Security and privacy~Vulnerability management}
\ccsdesc[300]{Software and its engineering~Software post-development issues}

\keywords{security updates, patch management, vulnerability management}

\maketitle

\section{Introduction} \label{section:introduction}

Timely patching of security vulnerabilities is vital for safeguarding software systems against cyberattacks \cite{mell2005creating, li2019keepers}. Failure in promptly applying security patches usually results in devastating consequences as a result of successful cyberattacks that particularly target software systems in mission-critical domains such as healthcare \cite{cishealthcarecyber, cyberattackhealthcare}. For example, a failure in the timely installation of an available patch exploited by the WannaCry ransomware cryptoworm brought down several critical systems of the National Health Services (NHS) in the UK \cite{wannacry}. Several studies \cite{mell2005creating, jenkins2020anyone} report that available but not installed security patches are one of the main reasons for successful cyberattacks, e.g., the Heartbleed security bug that remained unpatched in systems for several years \cite{hearbleed}. 

In order to understand the reasons for and provide solutions to the delays in applying security patches, an increased amount of research attention is being targeted at the software security patch management process, also called security patch management \cite{mell2005creating, li2019keepers, dissanayake2021software}. Software security patch management consists of several activities, such as detecting, obtaining, testing, installing and verifying security patches, and associated interdependent artefacts on which different types of stakeholders work \cite{dissanayake2021software}.

Like other areas of software engineering \cite{Automatedpatchase2020, turker2021efficient}, it has been claimed that automation has great potential for significantly improving the effectiveness and efficiency of security patch management to minimise the delays in applying available security patches \cite{araujo2020improving}. That has led to increased enthusiasm for devising and integrating automation support in different activities of the security patch management workflow. For example, automated approaches/tools for patch information retrieval \cite{al2005automated, rahman2013idispatcher}, vulnerability scanning \cite{bozorgi2010beyond, angelini2019mad}, assessment and prioritisation \cite{bozorgi2010beyond, kamongi2013vulcan}, patch testing \cite{dunagan2004towards, crameri2007staged}, patch deployment \cite{kashyap2016instant, araujo2020improving} and post-deployment patch verification \cite{prochazka2011race, chandra2011intrusion}. Despite these research efforts producing promising results for streamlining different activities of security patch management by providing automated solutions, delays in real-world security patch management still exist \cite{Accenturereport2021}. We assert that it is important to empirically understand the current status of automation, the real-world issues facing the current automation and the practitioners’ needs for improved automation support to develop suitable automated approaches and tools for effectively addressing delays in security patch management. To the best of our knowledge, there is little empirically known about these aspects of automation for security patch management.

Motivated by the need to empirically understand the role and user needs of automation for improving security patch management, we conducted an empirical investigation using in-depth semi-structured interviews with 17 practitioners from three organisations in the healthcare sector. We followed the Straussian version of the Grounded Theory method \cite{strauss1998basics} to systematically uncover the findings in a bottom-up approach. Our study findings contribute to the state-of-the-art understanding of research and practice by (i) providing an evidence-based understanding of the as-is state of automation in security patch management describing the current manual and automated tasks in the process, (ii) identifying the limitations of the current automation in practice, (iii) informing how automation support can be enhanced in effectively meeting practitioners' needs, (iv) explaining the role of the human in security patch management automation identifying the tasks and points in the process in which automation needs to facilitate a human-in-the-loop approach and (v) discussing the future avenues of research and providing a set of recommendations that can guide tool designers and researchers to address the identified gaps and user needs. 

\section{Background} \label{section:background}


Software security patches are \textit{``pieces of code developed to address security problems identified in software"} \cite{mell2005creating}. Organisations often prioritise security patches over non-security patches (e.g., feature patches) due to their criticality. Security patch management refers to the \textit{``multifaceted process of identifying, acquiring, testing, installing, and verifying security patches for the identified third-party vulnerabilities"} \cite{dissanayake2021software}. 
Existing studies \cite{li2019keepers, tiefenau2020security, dissanayake2021software} report that the process consists of five phases as described in Table~\ref{tab:patchmanagementprocess}.

\begin{table}[h]
  \caption{Phases of the security patch management process}
  \label{tab:patchmanagementprocess}
  \centering 
  \small
\setlength\tabcolsep{2pt}\def\arraystretch{1.5}\begin{tabular} {p{.12\textwidth} p{.34\textwidth}}
    \toprule
    {\sffamily \textbf{Process Phase}} & {\sffamily \textbf{Description of the Tasks Undertaken}} \\
    \midrule
    {\sffamily Patch information} \newline{} {\sffamily retrieval} & $\bullet$ {\sffamily Learning about new patch releases} \newline{} $\bullet$ {\sffamily Acquiring patches from third-party vendors and distributing to the target machines} \\
    \hline
    {\sffamily Vulnerability} \newline{} {\sffamily scanning,} \newline{} {\sffamily assessment and }\newline{} {\sffamily prioritisation} & $\bullet$ {\sffamily Scanning systems to identify existing vulnerabilities} \newline{} $\bullet$ {\sffamily Assessing the vulnerability risk based on organisational context} \newline{} $\bullet$ {\sffamily Prioritising patch decisions according to the vulnerability risk assessment} \\
    \hline
    {\sffamily Patch testing} & $\bullet$ {\sffamily Preparing for patch deployment by configuring the machines, handling patch dependencies and scheduling patch windows} \newline{} $\bullet$ {\sffamily Testing patches for inadvertent errors or side effects} \\
    \hline
    {\sffamily Patch} \newline{} deployment & $\bullet$ {\sffamily Installing patches on the machines} \\
    \hline
    {\sffamily Post-deployment patch verification} & $\bullet$ {\sffamily Verifying the patch deployment} \newline{} $\bullet$ {\sffamily Handling post-deployment issues} \\
    \bottomrule
  \end{tabular}
\end{table}

Although no work has specifically investigated the role of automation in the process in practice, the previous research has invested efforts aimed at providing automation for security patch management to bring efficiency into the patch management process and reduce delays \cite{dunagan2004towards, crameri2007staged, prochazka2011race, araujo2020improving}. Most of the previous efforts focus on automating specific tasks in security patch management, rather than supporting the end-to-end process \cite{dissanayake2021software}. A few studies have proposed solutions for automated patch information retrieval from multiple sources \cite{al2005automated, midtrapanon2019linux}, customised information filtering \cite{al2005automated, trabelsi2015mining}, information validation \cite{rahman2013idispatcher} and patch download and distribution \cite{rahman2013idispatcher}. For vulnerability scanning, a central platform integrating the scan results from multiple information sources has been suggested \cite{angelini2019mad, bozorgi2010beyond}. Towards vulnerability risk assessment and prioritisation, researchers \cite{kamongi2013vulcan, angelini2018vulnus, bozorgi2010beyond} have proposed customisable tools/approaches for vulnerability risk analysis in line with the industry standard, the Common Vulnerability Scoring System (CVSS) \cite{cvss}. 

Towards patch testing, a set of tools have attempted automated detection of faulty \cite{dunagan2004towards, crameri2007staged} and malicious patches \cite{kim2020new, kim2017study}. However, only limited attention has been given to recovering from crashes that result from faulty patches with minimum disruption. Several studies focus on automating patch deployment \cite{chang2005cross, midtrapanon2019linux}. To minimise downtime and service disruptions, several approaches have been developed; such as dynamic software updating (DSU) \cite{hicks2005dynamic}, JIT patching \cite{araujo2020improving} and instant kernel updates \cite{kashyap2016instant}. For post-deployment patch verification, automated approaches for verification of patch deployment \cite{prochazka2011race}, detection of exploits \cite{zhou2010always}, and repair of past exploits \cite{chandra2011intrusion} have been presented. An important observation is that only a few studies have rigorously evaluated the proposed solutions in real-world contexts \cite{dissanayake2021software, islam2022runtime}; an implication of little empirical understanding of how well the solutions have addressed the practitioners' needs.

Meanwhile, another set of studies has investigated the socio-technical aspects relating to security patch management, particularly the patch management process and its challenges \cite{li2019keepers, tiefenau2020security}, the role of coordination and collaboration \cite{nicastro2003security, huang2012patch, nappa2015attack, dissanayake2021grounded}, system administrators' practices, behaviour, and experiences \cite{crameri2007staged, dietrich2018investigating, li2019keepers, tiefenau2020security, jenkins2020anyone}, and reasons and mitigation strategies for delays in security patch management \cite{dissanayake2022cscw}. The understanding of the socio-technical aspects is considered important as security patch management is essentially a socio-technical endeavour whereby humans and technologies continuously interact for enabling team members to effectively coordinate and collaborate using the available tools \cite{dissanayake2021software}. An important observation is the highlighted need for human involvement in the patch management process despite the advancements in automation to reduce patch management delays  \cite{dunagan2004towards, crameri2007staged, maurer2012tachyon, angelini2019mad, li2019keepers, tiefenau2020security, dissanayake2021software}. However, why human involvement is needed in process automation and the desired balance between automation and human involvement has not been investigated.

Our work distinguishes itself from the above-mentioned works as it purports to provide a holistic understanding of the role of automation in security patch management grounded in evidence gathered from practitioners. We explain how practitioners have integrated automation into the security patch management process by describing the automated and manual tasks in the workflow, limitations of the current automation in meeting practitioners' needs, how automation support can be improved to assist practitioners, and the roles of humans in process automation, i.e., why and where is human involvement needed. Given the criticality of timely security patch management, this study's findings will provide a solid foundation to propose practical solutions that can address the limitations of current automation support for security patch management. 

\section{Research Method} \label{section:method}

This study was aimed at understanding the role of automation in security patch management in practice. We conducted a qualitative study using semi-structured interviews that are expected to enable us to gain a better understanding of real-world practices through practitioners' perspectives. To achieve this goal, we developed the following research questions (RQs):

\textbf{RQ1. What is the as-is state of automation in security patch management?} This RQ focuses on understanding the current state of automation in practice by identifying which security patch management tasks are currently supported by automation.

\textbf{RQ2. What are the limitations of current automation?} This RQ focuses on identifying the limitations of the current automation/tools to understand how well the current automation meets practitioners' needs. 

\textbf{RQ3. How automation in security patch management can be enhanced to support practitioners?} This RQ aims at understanding the kind of support practitioners expect from improved automation. The knowledge is important to design future tools to effectively meet practitioners' needs.  

\textbf{RQ4. What is the role of the human in security patch management automation?} This RQ focuses on understanding the roles humans play in security patch management automation and the reasons demanding human involvement. Such understanding can help in identifying the tasks for which future automation can be designed to facilitate a human-in-the-loop approach.

\subsection{Data Collection}

\textbf{Participants recruitment:} We collected data through semi-structured interviews with 17 practitioners from three organisations (Org A, B and C) in Australia. We used a mix of purposive \cite{schwandt1997sage} and convenience \cite{marshall1996sampling} sampling when selecting the organisations to ensure that our data are representative of the substantial area through which the findings emerge. 
Org A is a large health services agency responsible for maintaining the IT systems in an Australian state government’s healthcare sector. In addition to maintaining third-party vendors' software systems, they also maintained some medical software applications developed in-house. However, security patch management of operating systems (OS) and associated third-party applications were outsourced to Org B and Org C, which are IT services and consulting organisations. The non-security patches were handled by different teams in Org A, which is not included in our scope.

We requested the manager of Org A's security team to suggest potential interviewees engaged in security patch management in the case organisations. Then, we sent email invitations to the nominated candidates. The participant selection covered a diverse set of job roles in security patch management to achieve a rounded perspective. A pre-interview questionnaire was emailed to the participants to collect demographic information about them and their team. The demographics of the participants are presented in Table~\ref{tab:participantsdemographics}. The participants, on average, had 22 years of experience in security patch management. The participant (PID) and organisation identities have been omitted to preserve participant anonymity as per the human ethics guidelines. The last three columns refer to the number, types of machines or devices, and types of software components managed by the participants. 

\begin{table*}
  \caption{Participant demographics}
  \label{tab:participantsdemographics}
  \centering 
  \small
  \begin{tabular} {p{.02\textwidth}  p{.23\textwidth} p{.04\textwidth} p{.04\textwidth} p{.07\textwidth} p{.035\textwidth} p{.03\textwidth} p{.055\textwidth} p{.11\textwidth} p{.17\textwidth}}
    \toprule
    PID & Role & Yrs of \newline{Exp} & Org & Domain & Team & Team Size & No of \newline{Machines} & Machine Type & Software Component \newline{Type}\\
    \midrule
    P1 & Lead Solution Analyst & 20 & Org A & Healthcare & T1 & 65 & 250 & Server & Applications \anote{$^*$}, EOL \anote{$^+$} \\
    P2 & Team Lead & 25 & Org A & Healthcare & T2 & 17 & 3000 & Server & OS \\
    P3 & Infrastructure Specialist & 23 & Org A & Healthcare & T2 & 17 & 3000 & Server & OS, Applications, \newline{Drivers and Firmware} \\
    P4 & Server Administrator & 23 & Org A & Healthcare & T2 & 17 & 3000 & Server & OS, Applications, EOL \\
    P5 & Infrastructure Server Engineer & 17 & Org A & Healthcare & T2 & 17 & 3000 & Server & OS \\
    P6 & Manager (Infrastructure Server) & 25 & Org A & Healthcare & T2 & 17 & 3000 & Server & OS \\
    P7 & Service Transition Manager (EUC) & 35 & Org A & Healthcare & T3 & 3 & 35000 & Client & OS, Applications \\
    P8 & ICT Change Manager & 22 & Org A & Healthcare & T4 & 3 & 38000 & Server, Client, \newline{Network devices} & All Types \\
    P9 & Senior Manager (Security Services) & 20 & Org A & Healthcare & T5 & 6 & 38000 & Server, Client & OS, Applications \anote{$^*$}, EOL, \newline{Drivers and Firmware} \\
    P10 & Senior Security Advisor & 40 & Org A & Healthcare & T5 & 6 & 38000 & Server, Client & OS, Applications \anote{$^*$}, EOL, \newline{Drivers and Firmware} \\
    P11 & Manager (EMR) & 19 & Org A & Healthcare & T6 & 250 & 160 & Server & Applications \\
    P12 & Change Coordinator (EMR) & 19 & Org A & Healthcare & T6 & 250 & 160 & Server & OS, Applications, EOL \\
    P13 & Patching Lead & 3.5 & Org B & IT & T7 & 5 & 1800 & Server & OS, Applications \\
    P14 & Client Delivery Manager & 21 & Org B & IT & T8 & 50 & 160 & Server & OS \\
    P15 & Account Delivery Lead & 20 & Org B & IT & T8 & 50 & 2500 & Server & OS, Applications, EOL \\
    P16 & Account Run Lead (Operations) & 14 & Org C & IT & T9 & 130 & 38000 & Client & OS, Applications \\
    P17 & Senior Server Engineer & 25 & Org C & IT & T10 & 30 & 1500 & Server & OS \\
    \bottomrule
  \end{tabular}
  \smallskip
  \parbox[t]{\textwidth}{ $^*$~Includes both third-party applications and custom in-house applications. $^+$~EOL = End-Of-Life (Legacy) Software}
\end{table*}

\textbf{Semi-structured Interviews:} We conducted semi-structured interviews with 17 participants from 10 teams of three organisations from November-December 2021. The interviews lasted between 30 and 60 minutes and were conducted via Microsoft Teams because of Covid-19 restrictions. The first author led all the interviews, while the second and the third authors participated in seven interviews (author2 = 4 interviews and author3 = 3 interviews) along with the first author for helping in asking the follow-up questions and increasing the reliability of the interviewing process.  

The interviews were focused on security patch management of OS and third-party software applications in the context of healthcare. The open-ended interview questions covered the following areas: (a) role and responsibilities in the security patch management;
(b) the current status of the automation tools used in the process; 
(c) the challenges and limitations of the current automation tools;
(d) role of humans in security patch management automation and the reasons for human involvement, and
(e) the need for better automation support (i.e., the envisioned solutions and their features). 

All the authors jointly prepared the interview guide which was modified with mutual agreement when appropriate during data collection 
to reduce the potential bias. All the interviews were recorded and transcribed for analysis by the first author. The interviews were conducted in two phases in parallel with the data analysis. We continued with the data collection until the data analysis confirmed \textit{theoretical saturation} \cite{strauss1998basics}, for example, the last three interviews provided more examples for the emerged findings but no new categories or insights emerged.

\subsection{Data Analysis}

We used Grounded Theory (GT) \cite{strauss1998basics, glaser1967discovery} for data analysis as it aligned well with our study's goals of investigating a phenomenon in a real-world context, in this case, understanding the role of automation in security patch management in practice. We adopted Strauss and Corbin’s GT version (Straussian GT) \cite{strauss1998basics} as it provides a well-structured data analysis approach guided by open-ended and practice-based RQs \cite{strauss1998basics, stol2016grounded}. The data analysis procedure followed different types of coding: \textbf{\textit{open coding}}, \textbf{\textit{axial coding}} and \textbf{\textit{selective coding}} \cite{strauss1998basics}. The first author conducted the data analysis and created a Codebook containing all the codes and memos. The Codebook and interview transcripts were shared with all co-authors. The second and third authors cross-validated all the codes, concepts, categories, and core categories in the Codebook against the raw data (i.e., interview transcripts) to reduce bias and increase the reliability of the findings \cite{strauss2007basics}. The categories and their relationships were thoroughly discussed in weekly meetings and finalised after several rounds of revisions among all authors. Any disagreements were resolved in weekly detailed discussions between all authors throughout the process. The use of analytical tools such as diagramming and memoing facilitated the data analysis process. The interview transcripts were stored in NVivo, a qualitative analysis tool \cite{Nvivo}.

\begin{figure*}[h]
  \centering
  \includegraphics[scale=0.75]{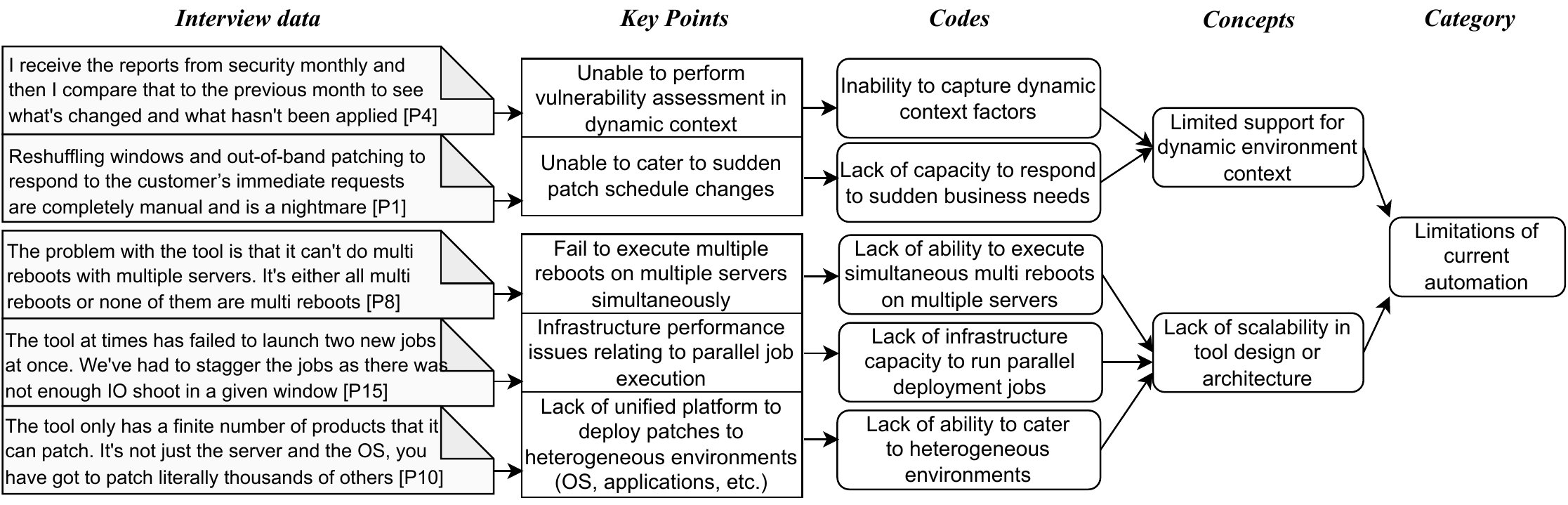}
  \caption{An example of the data analysis steps leading to a category from raw data as evidence.}
  \Description{Emergence of the category \textit{Limitations of current automation} from the underlying codes and concepts.}
  \label{fig:dataanalysiscoding}
\end{figure*}

During \textbf{\textit{open coding}}, we read through the interview transcripts line by line summarising the content as \textit{key points}. It was further encapsulated into \textit{codes} with short phrases. Constantly comparing the emerged codes within a single interview transcript and between different interview transcripts, gave rise to \textit{concepts} \cite{strauss1998basics}, a higher level of abstraction of the \textit{codes}. Similarly, we developed \textit{categories}, the next level of abstraction, as shown in Figure \ref{fig:dataanalysiscoding}. Next, we applied \textbf{\textit{axial coding}} in which we systematically linked the emerged categories at the level of their properties (i.e., ``characteristics of a category") and dimensions (i.e., ``variations within properties") to uncover the relationships among them  \cite{strauss1998basics}. This process of axial coding was guided through activities involving referring back and forth to \textit{memos}, i.e., notes written during open coding explaining the codes and their relationships, drawing the inter-relationships in diagrams, and refining them through frequent team meetings. The process continued until no new properties, dimensions, or relationships emerged for each category, which indicated \textit{theoretical saturation} \cite{strauss1998basics}. Finally, we applied \textbf{\textit{selective coding}} where we finalised the categories and integrated them with the main theme of the research, known as the \textit{central category} \cite{strauss1998basics}, in this case, the role of automation in security patch management.

The raw data obtained in this study cannot be shared because of confidentiality agreements with the industry collaborators, which happen to be in the health sector where data sensitivity has extra layers of governance. However, we made our interview guide containing the pre-interview questionnaire and interview questions publicly available~\footnote{\url{https://doi.org/10.5281/zenodo.6519830}}.  

\section{Results} \label{section:results}

The sub-sections of this section report our study's findings for different aspects of automation in security patch management. 

\subsection{As-Is State of Automation in Security Patch Management}

\begin{figure}[h]
  \centering
  \includegraphics[scale=0.7]{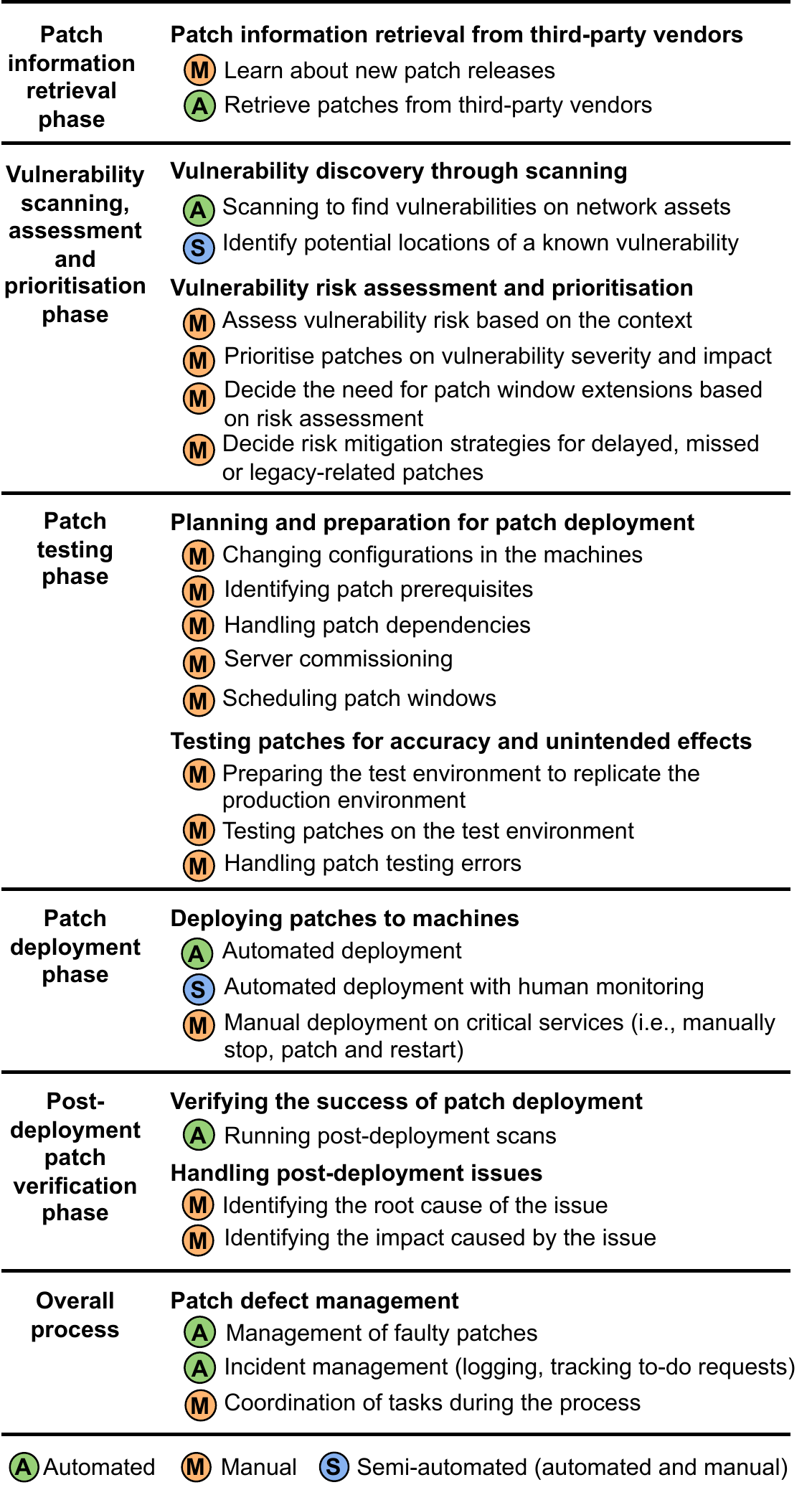}
  \caption{The as-is state of automation in each phase of the security patch management process. The \textbf{bold} text in the right column denotes the main tasks in each phase and the sub-tasks are listed underneath.}
  \Description{Summary of the as-is state of automation in security patch management.}
  \label{fig:asisstatefindings}
\end{figure}

Our findings to answer RQ1 reveal that whilst some tasks of security patch management have full or partial automation support, several of the tasks are carried out entirely manually. Figure \ref{fig:asisstatefindings} summarises the current state of the automation reported by the participants characterised by each process phase shown in Table~\ref{tab:patchmanagementprocess}. We describe how practitioners have integrated automation into the main tasks and the tools used below.

\subsubsection{Patch Information Retrieval from Third-Party Vendors} 

Most patch deployment tools automatically retrieve the patches from third-party vendor sites and distribute them to the patching towers. However, learning about new patch releases is a manual task whereby practitioners proactively search for patch release information using various sources such as direct vendor calls, security professional mailing lists, community Slack channels and online forums.

\hspace{2 mm}\textit{``The [vendor] schedules a call monthly for his premium customers. Then there's a security professionals' community where we share intelligence on incidents happening. For example, mid last year, a Microsoft patch caused problems on the printer and it was highlighted in that forum. So at least we had some forewarning to minimise the business impact."} - P9

\subsubsection{Vulnerability Discovery through Scans} 
Vulnerability scanning involves performing scans to locate the assets on the network and finding vulnerabilities on the assets which are automated using the tool ``Tenable.sc" \cite{tenable}. The practitioners schedule the vulnerability scanner (e.g., Nessus \cite{nessus}) to perform the scan and retrieve the results through automatically generated reports. However, identifying the potential locations of a known vulnerability sometimes requires manual effort; for example, writing a script to query what DLLs or java files the machine had in certain locations to find the locations of the Log4j vulnerability \cite{log4j}.

\hspace{2 mm}\textit{``We wrote a small batch file to query what DLLs or java files the machine had in certain locations and pushed out to all the servers and executed it without the tool. It created a results file locally on the machine and that was scraped and analysed later."} - P13

\subsubsection{Vulnerability Risk Assessment and Prioritisation} 
Based on the understanding of system vulnerabilities through the scans, practitioners manually perform the risk assessment and prioritisation to decide on patching. This is because the existing scanners fail to incorporate the organisational context and needs which are required for an accurate risk representation. Patch prioritisation is based on the vulnerability severity and impact. Then the need for patch window extensions is decided based on the risk assessment as some security patches can have a wide impact on several other services, for example, Log4j vulnerability \cite{log4j}. 

\subsubsection{Planning and Preparation for Patch Deployment} 
The practitioners engage in a set of tedious manual tasks involving changing configurations in the machines, identifying the patch prerequisites, handling patch dependencies, server commissioning (i.e., the process of building servers ready for deployment), and planning the patch schedules. In this, identifying patch prerequisites and scheduling tasks are perceived as the most challenging, necessitating significant manual overhead. 

Patch prerequisites (e.g., registry changes, group policy object (GPO) changes, preparation package installation) are identified by extensively reading through the patch release notes, knowledge base (KB) articles or information on the vendor's website. On the other hand, scheduling a patch window involves a series of tasks of extensive planning and organising with external stakeholders. Some participants use a third-party integrated risk management solution (e.g., Archer GRC \cite{glass}) for organisational governance and compliance with industry regulations; however, configuring the tool is a laborious task. The others manually schedule patches using an Excel spreadsheet. 

\hspace{2 mm}\textit{``Every month, I export the information of 1800 servers from the database to a large sheet with 27 columns and go through it to find the windows."} - P13

\subsubsection{Testing Patches for Accuracy and Unintended Effects} 
The participants perform testing in dedicated testing environments (e.g., ``Dev", ``Test" and ``Pre-prod") before rolling out the patches into production (``Prod"). However, due to resource constraints, client machine patching typically employs a staged deployment with a 10:30:60 rollout (i.e., systematically deploying patches to clusters of 10\%, 30\% and 60\% of machines without testing). Preparing the test environment to replicate the ``Prod" is also a laborious task due to shared access resulting in too many interdependencies. 

\hspace{2 mm}\textit{``If we find that the patch has an interdependency with an application, then my team has to work internally with the related teams to deploy the patch from Dev, Test, Pre-prod through to Prod which needs a lot of coordination."} - P11

\subsubsection{Deploying Patches to Machines} 
The participants use several third-party patch management software like Ivanti (server) \cite{ivanti}, SCCM (server) \cite{microsoftSCCM}, WSUS (server) \cite{wsus}, and VMware Workspace ONE (desktop) \cite{vmwareWorkspaceONE} for deploying patches based on their needs and budget. On average, 67\% of the servers (i.e., 1200/1800) are patched automatically. However, the success of automated deployment relies on the accuracy of server information fed into the tool and tool configuration. 

In an issue during automated deployment (e.g., the application not automatically restarting upon the server reboot), it is switched to semi-auto allowing practitioners to log into the machines to monitor the patching job while trying to investigate the issue. 
On average, 11\% of the servers (i.e., 200/1800) are patched semi-automatically. Meanwhile, 22\% of deployments, i.e., 400 of 1800 servers on average rely on decision-making that is extremely challenging or impossible to be automated; hence those systems are patched manually. For example, \textit{``we can't automatically shut down a system in the middle of an ongoing mission. So we tell the client we need to reboot the system and it'll be out for X time. They might say no, we've got four missions going on. So in such cases, there's no way automation is ever going to work there."} - P6

\subsubsection{Verifying the Success of Patch Deployment} 
The main verification technique used is scanning the systems post-deployment; the commonly used tools are Ivanti \cite{ivanti} and Tenable.sc \cite{tenable}. The participants reported that they target 99\% coverage of the server fleet as the remaining 1\% is usually never completed due to various reasons such as legacy systems. 

\subsubsection{Handling Post-Deployment Issues} 
This involves two manually performed activities: identifying what caused the issue and how wide the impact it has caused. 
Identifying the impact of the issue is important for deciding on the appropriate mitigation actions. The impact is initially measured in terms of the number of client complaints received and vendor announcements. The gathered information is then analysed to calculate the risk and identify the potential workarounds (e.g., rollback) to enable service continuity.

\hspace{2 mm}\textit{``It's a real pain in the bum to have patches fail because it can be quite intensive. We've had months where we had to roll back 20, 30, and 40 servers. And that's when it gets really ugly because the rollbacks are expensive like a full virtual server restore."} - P15

\subsubsection{Patch Defect Management} 
To manage the patch defects identified during patch testing through deployment to verification, participants use an in-house application life cycle management (ALM) solution. Incident management (i.e., logging and management of to-do patching requests) is supported by different tool choices such as third-party IT Service Management tools (e.g., Marval \cite{marval}) and in-house solutions. However, the coordination of tasks between teams is primarily done manually. 

\subsection{Limitations of Current Automation}

This section presents the limitations of the current automation/tools to answer RQ2. An understanding of the current solutions' limitations is important for identifying the areas of improvement for addressing the identified limitations as discussed in Section \ref{section:discussion}.

\subsubsection{Limited Support for Dynamic Environment Conditions}

Whilst many vulnerability management tools claim to have supported automated vulnerability assessment, they do not capture the dynamic context factors in vulnerability assessment leading to an inaccurate risk assessment. That is why practitioners are required to do the heavy lifting for risk assessment incorporating the organisational context. In addition, the current automation lacks adequate support for accommodating sudden changes in the schedules, for example, out-of-band (OOB) patching (i.e., outside the scheduled windows). Whilst many participants desire automation support to adapt to unforeseen changes, some believe that the coordination among different stakeholders and decision-making in such cases depend on human intuition that is difficult to be automated. 

\hspace{2 mm}\textit{``Automation, while great and saving time, also reduces the chance for special requirements, interaction and coordination with the business when things suddenly change. We have to work out what we can patch automatically and what still requires coordination with the business and all those things."} - P14

\subsubsection{Lack of Proper Support in Process Workflows}

Limited support for patch deployment preparation warrants significant effort for reading through release notes to identify patch prerequisites. 
Correspondingly, limited support for handling patch dependencies seems to be a common complaint. The lack of a holistic view of the system interdependencies requires practitioners to spend a significant amount of time and effort in identifying the dependencies during testing and troubleshooting deployment issues. The absence of automation support to handle legacy software dependencies creates further complications, often leading to delays in restarting services to avoid system breakdowns. Another drawback is the lack of support for detecting the need for multiple reboots. Although the tools are capable of automatically executing the scheduled reboots, identifying how many reboots are required is a manual task.

The current tools do not support identifying and remedying the service interruptions caused by incompatible dependencies during deployment. As a result, practitioners are forced to adopt effort-intensive workarounds to minimise resultant service disruptions. A lack of real-time report generation detailing the deployment errors leads to a lack of interpretability in verification leaving practitioners to spend hours troubleshooting the root cause of an issue. 

\hspace{2 mm}\textit{``Some tools are chosen because of the practicality, not because it's the evergreen solution. Sometimes automatically deploying a patch becomes impossible when the application does not restart correctly due to a broken dependency. So, we often do a very tedious manual process to stop, patch and restart the services."} - P2

\subsubsection{Lack of Accuracy of Output} 

Another deficiency is missing information (e.g., skipping some patches) in scanning leading to incorrect vulnerability reports. According to P4, discrepancies in the scan reports can be caused by a failure in running the scan completely or some information may be missing from the scan report. Inaccurate scan reports thus result in inaccurate vulnerability assessment leaving the system exposed to a myriad of attack opportunities. Lack of capacity to detect mid-cycle patch releases (i.e., superseding patches released during the patch cycle) is another limitation that leads to false positives in the scan reports.

\hspace{2 mm}\textit{``In cases of superseded patches, 
the tool reports them as missing patches. So we had to go through Microsoft's catalogue to find out that those patches have superseded the previous patches and notify the security team that this is what has happened."} - P5

\subsubsection{Lack of Scalability in Tool Design/Architecture} 

The lack of a unified platform to deploy patches to heterogeneous environments (i.e., multiple operating systems, dependent applications, etc.) is an important infrastructure limitation of existing tools. Microsoft SQL is a classic example that results in great annoyance for practitioners, forcing them to shift to manual deployment or juggle between several tools leading to often missing out on patches during deployment. 

\hspace{2 mm}\textit{``The tool has only a finite number of products that it can patch. Imagine the trouble when I have hundreds of products but it can only patch a half.
"} - P10

Another concern is the performance limitations in terms of the lack of capacity to run parallel deployment jobs and execute multiple reboots on different servers simultaneously. These limitations demand practitioners to spend a lot of effort in careful planning to cope with the time and service availability constraints.  

\hspace{2 mm} \textit{``The tool at times has failed to launch two new jobs at once so we've had to stagger the jobs. In other words, there wasn't enough IO shoot enough jobs out in a given window."} - P15


\subsubsection{Service Disruptions During Patch Deployment}

A key limitation of the current automation is the service downtime resulting from reboots forcing practitioners to rely on workarounds like server clustering and failovers. Despite the constant struggle to minimise service disruptions, the workarounds also require extensive effort for planning and execution. Similarly, the necessity to do multiple reboots in some cases is a frequent frustration given the narrow patch windows and extended service interruptions.  

\hspace{2 mm} \textit{``It's very challenging as the application is unable to support redundancy or high availability. If you take one server down then the technology should be able to continue to run."} - P2

\subsubsection{Lack of Usability}

The notifications indicating an error in the current patch deployment tools do not suffice for the level of understanding required for participants to detect errors. The lack of meaningful error messages leads to insufficient information for troubleshooting deployment failures causing practitioners to spend significant time and manual effort ``\textit{finding where to look for what}" (P10).

\hspace{2 mm} \textit{``It's scary when something breaks after patching because no one knows where to start looking. So someone has to look at the error messages to understand what does it mean. 
How do you automate something if you don't know what you're looking for?"} - P10

\subsection{Practitioners' Needs for Enhanced Automation} 

This section answers RQ3, the desires expressed by the participants in response to the interview question \textit{``what tasks do you wish would have been better supported by the technology and how?".} We raised this question to better understand what practitioners really need to fill the gaps in the current automation to effectively meet their needs. 

\subsubsection{Automation Support for Patch Information Management}

The participants desire a single platform for retrieving trusted patch information from multiple sources covering new patch releases, mid-cycle releases, delays in patch releases, and potential patch adverse effects. Such a platform would assist them in making informed decisions about the patch application, the level of testing required and finding workarounds for potential adverse effects. 


Additionally, the participants wish that a system provides an analysis of the potential impact of new patches based on the patch information retrieved from external sources. P3 recalled a situation wherein they spent significant time searching through public forums to find out about a faulty patch release.

\hspace{2 mm} \textit{``There was a Windows OS security patch rolled out this week that prevented Windows 10 desktops from being able to print on Windows 2003. The patch didn't say exactly what it was doing, just that it's a security patch, but behind the scenes, it has changed the protocol."} - P3

\subsubsection{Central Platform Integrating Vulnerability Scanning and Risk Assessment}

The participants reported the need for enhanced filtering and customisation on vulnerability scanning that enables better filtering and information sorting to identify outstanding vulnerabilities at a glance. Additionally, the compiled wish list includes the ability to easily search through a myriad of vulnerabilities to track the remediation status and generate scan reports with better visualisations as the current report only provides summary statistics in Excel spreadsheet format. Another need is the capability to integrate vulnerability scanning and assessment reports with other tools such as the configuration management database (CMDB). While some existing platforms (e.g., Archer GRC Solution \cite{glass}) enable practitioners to view the reports generated through the vulnerability scanner, it requires a lot of manual effort on the customisations. 

\hspace{2 mm} \textit{``If the tool could communicate directly to [the scanner], then we can do some better filtering and information sorting. And if it can be linked to our internal knowledge bases, no need for manually tracking or storing with any individuals."} - P4

\subsubsection{Automated Preparation for Patch Deployment}

The need for automation support for identifying the patch prerequisites was highlighted by many participants. As described in Section 4.1.4, detecting prerequisites is currently a manual effort of reading through the release notes. However, the execution of prerequisites is considered difficult to be automated as selecting the optimum configuration needs reasoning based on the environment, therefore demanding human intervention.

\hspace{2 mm} \textit{``The tools do not consider what vendors have written in their release notes. The release note may say, to complete this patching, you also need to set up this configuration in this manner, and a lot of times this is missed. So a system that can flag which patches require manual intervention to adjust settings would come in handy."} - P9


\subsubsection{Automation Support to Articulate Patch Scheduling}

Another noteworthy desire from many participants was automation support for articulation work in patch scheduling as it involves a series of cooperative tasks (as described in section 4.1.4). The existing tools lack support in managing the integral set of patch scheduling tasks. The participants desire a single dashboard view of all patch schedules enabling easy identification of patch windows based on availability, modification of schedules, tracking status and schedule changes, interactive communication between stakeholders, and integration with the patching tool to export the schedules straight to deployment. Such a platform would assist in articulating patch scheduling and rescheduling tasks among distributed stakeholders to speed up patch deployment. 

\subsubsection{Automated Patch Deployment With Better User Control}

Whilst the current tools are capable of automated patch deployment, the participants expressed the need for enhanced automation with more user control in the capabilities described below. Reduced or no system downtime to address the challenges of service disruptions during reboots and exhaustive manual overhead of the stop-patch-restart process described in Section 4.2.2.  

\hspace{2 mm} \textit{``It would help a lot if there's a way of figuring out non-rebooting patches so that we could patch and keep going and it wouldn't matter when you patch as much."} - P8

The capability to automatically execute simultaneous multiple reboots on multiple servers is preferred as the current manual activity often leads to missing them leaving practitioners with ``no way to catch up" due to time and resource constraints. Another need is a unified platform capable of supporting patch deployment across heterogeneous environments, particularly beneficial in alleviating obstacles associated with shared access in an environment. In addition, a few participants indicated the need for improved usability of patch deployment tools with increased efficiency. 


\subsubsection{Automated Patch Deployment Verification And Recovery}

Several participants expressed the need for improved automation for verifying patch deployment and detecting post-deployment issues, particularly reporting. This need was emphasised by many as the current verification tasks require significant manual overhead thus often leaving them neglected. Real-time reporting is preferred by the participants over monthly summary reports as it enables detecting and responding to issues promptly. Additionally, providing meaningful error messages to aid the troubleshooting and automated recovery of patch deployment failures is also desired.  

\hspace{2 mm} \textit{``Verification of the deployment would be the biggest plus, so we wouldn't find an issue a week later. And then better reporting on patching success and the exceptions. A summary at the end of the month to me is a waste of time, I want as close to real-time."} - P10

\subsubsection{Improved Configuration Management Database With an Overview of System Interdependencies}

A configuration management database (CMDB) displays the configuration items (e.g., server, application, router, etc.) in a managed environment and how they interact with each other. As mentioned by P1, \textit{``a good CMDB will have whatever you want in there, whereas at best we have got a list of servers that may or may not be up to date with no relationships between them. It is something that helps you make informed decisions, especially around change management, be it patching or otherwise. I think that is an important vector into patching and the culture of how stressful it makes people".} The participants indicated the need for an improved CMDB providing a real-time overview of the system's patch state including information about the server status, patched date, decisions (e.g., server exemptions), and better filtering.


A single view of system interdependencies would be helpful to reduce the manual overhead of handling issues, particularly legacy software dependencies wherein the current process relies on human knowledge and expertise. It also facilitates the understanding and coordination of manual patching and patch scheduling between teams, thereby reducing the delays and risks of additional outages to critical services. A few expressed their desire for predictive analysis on the impact of patch dependencies, for example, predicting the list of interdependencies that could potentially be impacted by a particular patch to guide practitioners in patch testing. 

\subsection{Role of Human in Process Automation}


This section answers RQ4 by describing the crucial roles of humans in the security patch management process automation and explaining why and where human involvement is needed. 

\subsubsection{Gain Control Over Uncertain and Dynamic Environment Conditions}

The evolving conditions of the environment resulting in unpredictable changes to schedules are one of the main reasons demanding human involvement in automation. In such cases, the participants revealed switching to manual patch deployment to obtain increased control of the situation. Below are some \textit{use cases} that describe the roles humans play in a dynamic context.

\textit{Emergency patches} released to fix critical vulnerabilities require urgent attention as the patches need to be deployed within 48 hours of release. The unanticipated event and urgency of the task calls for human involvement in careful planning and execution in an OOB window, which is not possible through automation.

\textit{Lack of visibility into patch load dynamics} during the patch window (e.g., how many patches get applied, how long it takes to apply) is another reason that requires humans to be involved in instructing the tool of the subsequent actions (e.g., revert the installed patches, extend the window by x hours) to maintain system availability.

\textit{Unforeseen errors in patch deployment} resulting from faulty patches create the need for understanding the severity and impact of the situation, wherein humans are accountable for service continuity. Despite the uncertainty, incompatible dependencies exacerbate the challenges demanding an increased level of human involvement to resolve the broken dependencies. This is because new patch releases only consider the compatibility with the most recent vendor-supported software versions.

The \textit{dynamic environment context calls for a high degree of coordination} to manage the interrelated tasks between stakeholders. In mission-critical contexts, some coordinating tasks such as safety checks (e.g., time to deploy the patch) are extremely difficult to automate. Many participants do not believe that coordination tasks such as finding an outage window or negotiating a patch window extension can be automated as expected.

\hspace{2 mm}\textit{``From my point of view, the biggest problem is coordination and that's a purely human-driven process. I can’t see how the AI or a machine with various tools can help us find an outage window."} - P11

\subsubsection{Contextual Awareness-Based Decision-Making}

Increased complexity of tasks together with existing tools' lack of domain expertise and inflexibility demand human involvement in the appropriate decision-making. Humans’ contextual awareness, which cannot be fully automated away, is key for making the right decisions during uncertain and complex events. For example, \textit{shared services} necessitate the need to handle a large number of interdependencies in managed systems. Certain interdependencies such as legacy software dependencies entail significant domain expertise in assessing the risks of service interruptions. 


\hspace{2 mm}\textit{``One of the biggest things is understanding and trying to work out all the combinations of interdependencies. If we patch this one, will it break the other one or do we have to upgrade this one and this one? That is a big challenge as the more applications you put on one server, the more possibilities of interactions, hence more combinations that you've got to consider when patching."} - P10

\textit{High severity and impact of services} exacerbate the challenge of service downtime resulting from reboots. The criticality of service downtime requires an increased sense of agency in initiating, executing and controlling the patch deployment, where human sense-making is needed to decide on the right action. 

Further, the large volume of machines with myriad software versions, machine clusters and patch levels result in \textit{too many configuration options} leading to cognitive overload. The patch exemption requests from clients further complicate the issue. Hence, it necessitates human involvement in selecting and implementing the suitable configurations as the automation is not capable and trusted in reasoning, classifying or predicting the configuration options. Other example scenarios include making informed decisions about the need and number of extra reboots during a patch window and implementing workarounds to maintain system availability during post-deployment errors based on the severity and impact. 

\hspace{2 mm}\textit{``Very often human decision-making is needed. That extra setting really depends on your environment. And to see how your environment may be relevant, you have to assess how the risk is applied to your context."} - P9

\subsubsection{Handle Legacy Systems In Place.} 

Having \textit{unsupported legacy systems in place} is one of the fundamental reasons requiring human involvement. While the unsupported software poses a huge threat leaving several attack vectors open for exploits, many organisations retain legacy systems because of the service criticality and high cost and complexity involved in migrating legacy systems. Lack of an upgrade path for legacy applications and lack of support from vendors place the burden of handling incompatible legacy software dependencies on the practitioners' shoulders alone. 

\hspace{2 mm}\textit{``When we have problems with legacy systems, we find someone who knows the system to get it working again and then we don't touch it. If we can find someone who knows to fix it, beauty but if we don't, we wouldn't know what to do."} - P14

Another instance is \textit{executing multiple reboots}. The legacy software requires multiple reboots for getting the software up to date. As the number of reboots required depends on the system being patched, a human understanding of the context is needed to identify the exact number of reboots. Further, since multiple reboots can potentially exceed the patch window, human involvement is needed to take remediation action depending on the conditions.  

\subsubsection{Adapt to the Organisational Needs and Culture}

Organisational factors including the culture, policies and needs play an important role in the need for human association in process automation. This finding provides additional evidence to previous work \cite{li2019keepers, dietrich2018investigating}, which also reported the influence of the organisation’s internal policies and management on system administrators in the process.

\textit{Handling negative perceptions about security} in the organisational culture, for example, lack of interest, freedom and priority for security, result in poor practices or decisions that adversely impact security patch management. For example, non-security teams tend to neglect security patching, higher management delays patch approval decisions, and stakeholders refuse to corporate. In such cases, human involvement is essential in getting people to understand the need for security patching and negotiating the patch schedules maintaining a balance between the need to patch and maintaining system availability. Another instance is the \textit{need for managing resistance to change}. Interestingly, in some cases like patch scheduling, some participants reported they are resistant to shifting to an automated solution as they have got used to manually doing it with Excel spreadsheets. 


\section{Discussion and Future Work} \label{section:discussion}

We discuss the findings from this study and its broader implications for practitioners and researchers to derive some actionable insights and identify the areas for future efforts aimed at providing automation solutions for security patch management. \\

We present an evidence-based understanding of the role of automation in security patch management that describes the as-is state of automation in practice, its limitations, practitioners' needs for improved automation support and the role of the human in \nobreak tsecurity patch management automation. The evidence from Section 4.1 shows that a majority of the tasks in the security patch management process are performed manually (see Figure \ref{fig:asisstatefindings}). 
We find several reasons for the current state of manual work in the process. The key reason is the limitations of the current automation support reported in Section 4.2. This situation stems from different factors such as limitations in the existing tool capabilities (e.g., capturing organisational context in vulnerability assessment), lack of specific features (e.g., identifying prerequisites from patch release notes), performance limitations (e.g., inability to execute parallel deployment jobs), infrastructure constraints (e.g., lack of ability to deploy patches to heterogeneous environments), and usability limitations (e.g., lack of meaningful error messages). These limitations show the gap areas that present excellent opportunities for researchers and tool builders for providing advanced automation/tooling support.

Further, the evidence from Sections 4.2 and 4.3 indicates that certain limitations of the current automation solutions require the practitioners to perform several tasks manually which usually causes delays in patching. We argue that the tool design may have failed to take into consideration why and how automation is actually used in practice. We anticipate our findings would provide a solid understanding of what enhancements in automation support are needed (Sections 4.2 and 4.3) and how those enhancements will be used in practice (Sections 4.1). Such understanding will also be beneficial to shaping future work to effectively address the practical concerns in security patch management. The findings of Section 4.3 can provide an opportunity for tool builders to identify the features that would enable their tools to add more value and practical utility in practice. We recognise that these findings might not capture all the desired tool features as the participants might not be aware of other available tools and their capabilities. This limitation also prompts future work for a mapping of the features of the existing patch management tools to the practitioners' desired needs to scope down \textit{what is missing where}. 

Given our findings are context-dependent, i.e., in the healthcare domain, future research is needed for broader validation in different contexts and using additional data sources (e.g., large-scale surveys). The findings from this study can be extended to other domains to identify additional desired features with concrete requirements. Similar to previous work \cite{tiefenau2020security, dissanayake2021software}, we foresee the value of future work to evaluate the tools in real-world contexts to better understand how well the tool meets the industry needs.

Another set of reasons for the increased manual work stems from the socio-technical implications of human and machine interaction in the security patch management process (reported in Section 2). These reasons demand an in-depth investigation of the critical roles that humans play in gaining control of uncertain situations, understanding the context, making sense of the available information and accordingly making informed decisions as described in Section 4.4. Our findings identify the tasks and points in the security patch management process that require a balance between human control and automation so that future automation can be designed to facilitate a collaborative relationship with humans. The findings present opportunities for future research in \textit{``Human-AI collaboration"} \cite{dellermann2019future}, an emerging research paradigm which combines human and machine intelligence to collectively achieve a goal. We also suggest further research to investigate the challenges in developing patch management tools that harness collective human and machine intelligence. 
Based on our findings and the existing research on ``Human-AI collaboration" \cite{kamar2016directions,  amershi2019guidelines, crowderHAI}, we propose some recommendations that can guide future tool development to address the limitations of the current automation solutions. \\


\textbf{An Integrated Platform Offering Support Across All Process Phases.} The lack of an end-to-end automated solution that supports all phases of the security patch management process has been an important anti-pattern recognised in our study, also reported by another study \cite{dissanayake2021software}. This situation forces practitioners to use multiple tools for different tasks in diverse environments resulting in increased errors and delays. Since building a unified solution capable of supporting heterogeneous environments (e.g., Windows, Linux, Mac) is highly complex due to the inherent differences in each environment, there is a need for a consolidated platform that can offer end-to-end patch management life cycle support in a single environment to address a majority of the limitations of the existing solutions. Moreover, such a consolidated environment will also benefit from being interoperable with external third-party solutions to cater for the needs of a multi-vendor environment. \\


\textbf{Human-Machine Collaboration for Patch Management.} We envision the integration of human and machine intelligence can be beneficial in several security patch management tasks reported in Section 4.4. For patch prerequisites identification, we envision a system that not only leverages combined human and machine intelligence but also learns from each of the collaborative partners (i.e., human and machine) to improve future prediction accuracy. Specifically, a machine can convert the unstructured data in patch release notes to a structured format and identify the prerequisites, and provide the summary results to the human partner to ease the selection of the optimum configuration options based on domain knowledge. Further improvements can be made by leveraging machine learning techniques to predict the most suitable configuration options that can guide practitioners to make quick decisions.

Similarly, joint human and machine intelligence can benefit in identifying and handling machine configurations and patch dependencies. A machine can identify the interconnections between configurations and patch dependencies, and predict the potential dependency breakdowns to reduce the possibilities of human errors in handling them and the expected delays in manual work. Another task that can benefit from human and machine collaboration is patch deployment. When there are errors during patch deployment, a machine can inform the deployment failure to all dependent stakeholders and guide them about the potential subsequent actions (e.g., revert deployment, estimates of patch window extensions). \\


\textbf{Human-Centred AI Explanations to Assist Contextual Decision-Making.} To support human decision-making based on contextual understanding (Section 4.4.2), we propose that future automation focuses on enabling human-centred explanations. As suggested by the Human-AI Interaction design guidelines \cite{amershi2019guidelines} ``make clear why the system did what it did", we anticipate such a capability will assist humans in understanding a machine's predictions thereby allowing them to make better decisions based on the context. For example, in selecting the optimum configuration during patch testing, a machine can provide the basis for the prediction. As another important design functionally, such a system shall enable interactive bi-directional communication. Interactive explanations enable humans to interact with machines \cite{lai2020harnessing}; for example, by editing input and changing the prediction based on the context. Further, a machine can learn from human interactions and evolve to improve its predictions by leveraging machine learning training techniques such as Reinforcement Learning and Active Learning. \\


\textbf{Decision Support for Patch Scheduling.} The evidence reported in Section 4.1.4 shows that sudden changes to the patch schedules demand an arduous manual task of patch rescheduling. Further, it can produce a cascading effect on the dependent patch schedules resulting in increased complexity for rescheduling and further delays in patching. Responding to these challenges, we suggest developing a smart decision support system to guide practitioners toward accurate and quick decisions about patch scheduling and rescheduling. The envisioned system should be able to identify the cascading effect on dependent patch schedules, i.e., which schedules will likely get affected and how much is the impact, based on the identified patch dependencies and prerequisites. Accordingly, the system should provide an estimate of the patching delay. We argue that these insights if presented in simple visualisations (e.g., graphs), would lead practitioners to make informed decisions faster about rescheduling.


\section{Limitations} \label{section:limitations}

In this section, we discuss the potential threats to validity and how they were mitigated based on the guidelines reported in \cite{runeson2009guidelines, maxwell1992understanding}. 

\textbf{External Validity: } Our findings do not claim to be generalised from a sample to a population but rather applicable to the studied cases like most qualitative research. Our findings are limited to the practitioners in the case organisations in the domain of healthcare studied in-depth to provide a holistic understanding of the studied topic grounded in evidence that is not attainable through broader but shallower approaches (e.g. surveys). 
However, we believe that our findings can be recreated and adapted in other similar contexts; for such purpose, we have provided sufficient details about the studied participants and the research methodology (section 3) to facilitate transferability.

Concerning data representativeness, we are aware that our data collection is limited to the interviews although it covers a wide range of the participants' roles and experiences. We acknowledge that data triangulation, e.g., additional cases, observations or surveys, will be useful for future studies to verify this study's findings and also extend the scope.

\textbf{Construct Validity: } To mitigate this validity threat, all authors collectively prepared the interview guide and pre-interview questionnaire. 
The interview questions were revised through several iterations following a pilot interview which was held between authors (i.e., first author as the interviewee, third author as the interviewer, and others as observers). It assisted in finalising the questions' scope, clarity and duration. However, as the findings were not verified with the participants through member checking, there is a threat of potential misinterpretations from the interviews.

\textbf{Internal Validity: } To alleviate potential internal validity threats, we included the participants who are involved in security patch management and ensured a representation covering all aspects of security patching in the case organisations, i.e., server and desktop. Further, the participants had 22 years of experience on average, mitigating the risk of the participants’ lack of expertise.

\textbf{Reliability: } To ensure the reliability of the findings, the data collection, analysis process, and the emerged findings were thoroughly discussed among all authors in weekly meetings, as described in Section 3.2. Furthermore, two researchers were present in seven interviews to minimise the threat of researcher error.

\section{Conclusion} \label{section:conclusion}

In this paper, we report an empirical study aimed at providing an evidence-based holistic understanding of the role of automation in security patch management. Based on semi-structured interviews of 17 practitioners, we conclude whilst the security patch management workflow in practice incorporates manual and automated approaches, the majority of the tasks are performed entirely manually due to the limitations of the existing automation/tools and the lack of capacity of the available automation to handle certain use cases demanding human involvement. Based on an improved understanding of the practitioners' perceived needs for enhanced automation support, this study enables us to explain why and where human involvement is needed in security patch management automation. Our findings indicate that human involvement in certain tasks in the workflow, particularly concerning contextual awareness-based decision-making and human intuition, is crucial in security patch management. That is why we propose that future patch management tool development should aim at supporting human-machine collaboration by leveraging the best of both capabilities. For future work, we outline how and what functionalities may be needed to effectively address the current gaps and user needs whilst embracing the desired balance between automation and human control in the security patch management process. 

\begin{acks}
The authors thank our industry collaborators, without whose help this research would not have been made possible. We sincerely thank all the participants who generously shared their time and experiences in interviews with us. This study was conducted under the University of Adelaide Human Research Ethics Committee Application ID H-2020-035.
\end{acks}

\bibliographystyle{ACM-Reference-Format}
\bibliography{bibliography}


\end{document}